\documentclass[11pt]{article}

\usepackage{lipsum} 

\usepackage{bm}
\usepackage{floatrow} 
\pdfoutput=1
\usepackage{amsmath,amsfonts,amsthm}
\usepackage{esdiff}  
\usepackage{booktabs}  
\usepackage{url}  
\usepackage{hyperref}  

\usepackage[tableposition=bottom]{caption} 

\usepackage{cleveref}  
	\crefname{equation}{equation}{equations}
	\crefname{figure}{figure}{figures}	
	\crefname{table}{table}{tables}
\usepackage[skip=1.5pt,font=small]{caption}

\usepackage{bbm}

\usepackage[usenames,dvipsnames,svgnames,table]{xcolor}

\usepackage{graphicx}

\usepackage{tikz}  

\usepackage[sc]{mathpazo} 
\usepackage[T1]{fontenc} 
\usepackage{microtype} 

\usepackage{multicol} 

\usepackage[margin={1cm,1.5cm}]{geometry}
\usepackage{booktabs} 
\usepackage{float} 
\usepackage{hyperref} 

\usepackage{lettrine} 
\usepackage{paralist} 

\usepackage{abstract} 

\usepackage{titlesec} 
\renewcommand\thesection{\Roman{section}} 
\renewcommand\thesubsection{\Alph{subsection}} 
\titleformat{\section}[block]{\large\scshape\centering\bfseries}{\thesection.}{1em}{} 

\titleformat{\subsection}[block]{\scshape\centering}{\thesubsection.}{1em}{} 

\usepackage{fancyhdr} 
\DeclareCaptionFormat{myformat}{#1#2#3\hrulefill}
\usepackage[T1]{fontenc}
\usepackage[utf8]{inputenc}

\usepackage[parfill]{parskip}
\usepackage{lipsum}
\usepackage{authblk}
\makeatletter
\title{\vspace{-15mm}\fontsize{13pt}{13pt}\selectfont\textbf{My cat Chester's dynamical systems analysyyyyy7777777777777777y7is of the laser pointer and the red dot on the wall: correlation, causation, or SARS-Cov-2 hallucination?}} %
\author[1,2]{Eve Armstrong\thanks{evearmstrong.physics@gmail.com}\thanks{https://reality-aside.com/aprilfool}\thanks{\textit{First authorship is under dispute by the allegedly-second author.}}}
\author[ ]{Chester}
\affil[1]{Department of Physics, New York Institute of Technology, New York, NY 10023, USA}
\affil[2]{Department of Astrophysics, American Museum of Natural History, New York, NY 10024, USA}\par
\date{(Dated: April 1, 2021)}
\begin{document}
\maketitle 
\begin{abstract}
\noindent
My cat Chester investigates the elusive relationship between the appearance in my hand of a silver laser pointer and that of a red dot on the wall, or on the floor, or on any other object that resides within the vicinity of the laser pointer.  Chester first assesses preliminary establishments for causality, including mutual information, temporal precedence, and control for third variables.  These assessments are all inconclusive for various reasons.  In particular, mutual information fails to illuminate the problem due to a dearth of information regarding what the laser pointer might have been doing at times following Chester's first awareness of the dot.  Next Chester performs a formal reconstruction of phase space via time-delay embedding, to unfold the gggggggggggfffgfgtredvteometry ,mmmm.........,.,,......,.mmmmmmmmmmmmmmmmmmmmmmmmmmmmmmmmmmmmmmmmmmm of the underlying dynamical system giving rise to the red dot's trajectory.  The resulting attractor does not resemble a laser pointer.  The reconstruction could, however, be flawed, for example, due to the short temporal duration of the dot's observed trajectory.  Finally, the red dot could be a hallucination: a symptom brought on by COVID-19 - because, well, these days pretty much anything might be a symptom brought on by COVID-19.  On this note, Chester's kitten brother Mad Dog Lapynski offers an independent check on the red dot's existence.  Moreover, the results of this study are inconclusive and ca[pokilki[[[[[ll for follow-up.
\end{abstract}

\section{INTRODUCTION}
\begin{multicols}{2}

\begin{figure}[H]
\centering
\includegraphics[width=0.7\textwidth]{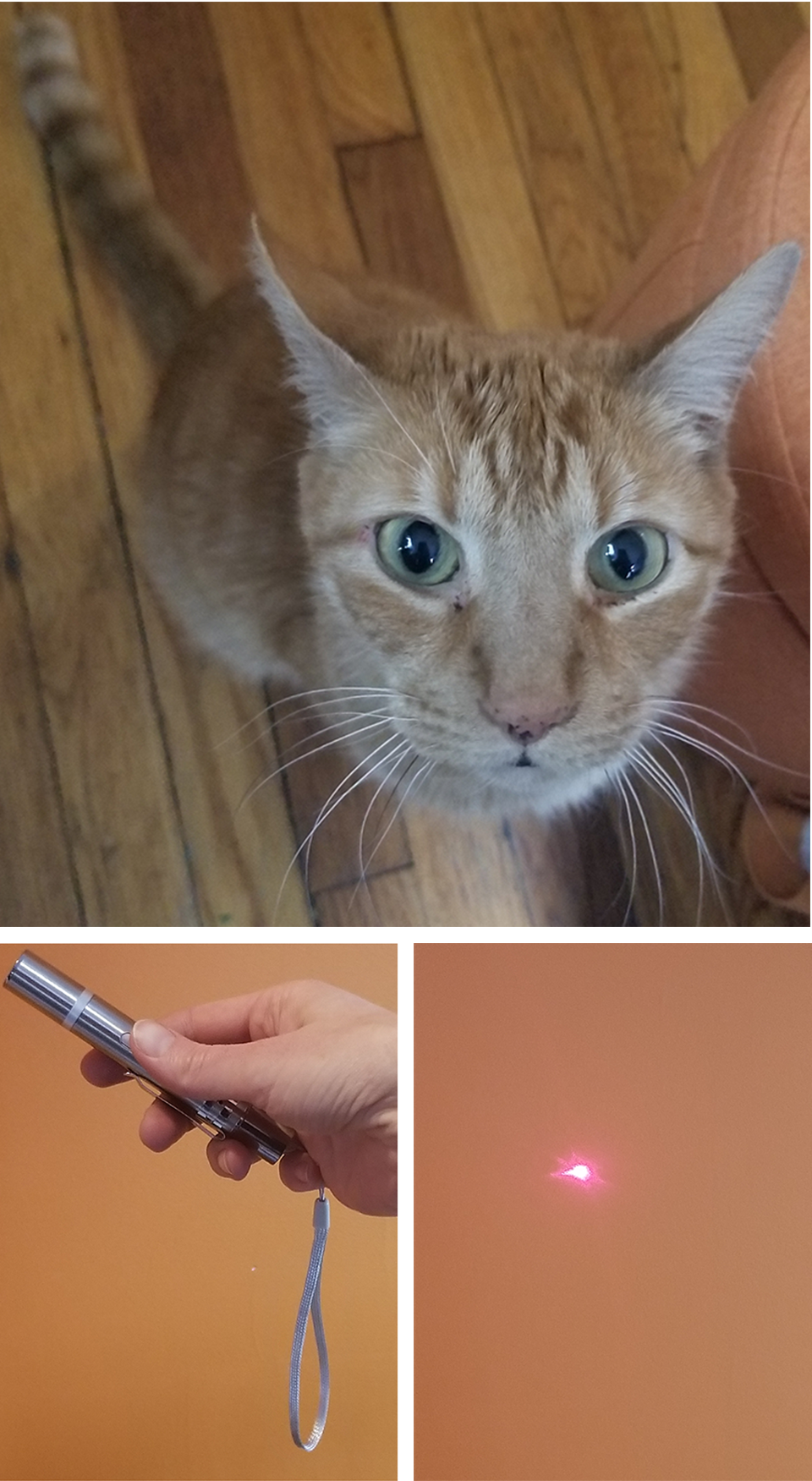}
\caption{\textbf{Chester, laser pointer, and red dot.}}
\label{fig:fig1}
\end{figure}

Chester (Fig.~\ref{fig:fig1}, top panel), an orange tabby, is a cuddling, clawing, purring, biting, trilling, hissing, head-bumping bipolar ball of fire.  In one moment he lovingly purrs underneath a head scratch, and in the next attempts to amputate my hand as punishment for having head-scratched just a wee bit too long.  He weighs eight pounds, six hundred\footnote{This is the typical weight of a Barbary lion~\cite{lion}, the largest in the world.} of which is ego.  In truth, Chester's endearing~\cite{vet1} and ferocious~\cite{vet2} personal attributes are too intertwined to decouple in writing. 

Among his diverse quirks, it is Chester's canniness that is the fodder for this paper.  Specifically: Chester possesses a distinctly \textit{un}canny talent for predicting events within his observable Universe.  He is poised statuesque on the kitchen counter by the time I enter to scoop out 2.5 ounces of Purrlicious Fishes.  He is peering wide-eyed out the livingroom window before our neighbor emerges to take her leashed parrot for a daily fly.  And he always head-bumps me one to five minutes prior to my morning alarm.  No computational pattern recognition algorithm holds a candle to Chester's fuzzy orange head.

And Chester flexes this predictive muscle not only on the simple case of one variable's cyclic behavior.  He also probes multivariate scenarios.  Specifically, those involving correlations.  The term 'correlation' brands any kind of statistical relationship between two or more variables.  For example, upon beholding a silver laser pointer in my hand (Fig.~\ref{fig:fig1}, bottom left), Chester will scan the space nearby until he locates a red dot (Fig.~\ref{fig:fig1}, bottom right).  

While Chester suffers no scarcity of things to chase, he is particularly taken with this red dot.  What renders the dot more intriguing than other chaseable objects is that - to date - he has \textit{never been able to catch it}.

Indeed, this dot is no ordinary dot.  To put it mildly, the dot displays unusual physical properties.  It is not, for example, a speck on a sweater that can be flicked or pinched or brushed away.  On the contrary, this dot seems to lack any material substance whatsoever.  It flits from surface to surface without once appearing in the air between the surfaces.  It vanishes and reappears in the blink of an eye, seemingly at the speed of light.  The laws of physics as we know them forbid this: no matter particle can travel at light speed~\cite{schutz2009first}.  Moreover, this dot truly appears to be not of this world.

Small wonder, then, is Chester's obsession.  What is this incorporeal dot?  Is it an ethereal spirit from a netherland?~\cite{broch2000yellow,james2007collected,casper,freddy} The tooth fairy~\cite{armstrong2012non}?  Batman~\cite{batman}?  And whence came it?  Regarding this last question, Chester may have a clue: the laser pointer correlation.

Now, while a correlation between two variables is rather straightforward to identify, digging at the nature of the underlying relationship is a slippery problem.  This slipperiness is not appreciated by all.  Some are fond of assuming that a correlation necessarily implies a specific causal relationship, especially if it concerns telomeres~\cite{sleepAge,internallyFit}, a full moon~\cite{moonBabies,moonCrime}, or sex drive~\cite{sexBananas,sexSwim,sexZodiac,sexMusic,sexPurple,sexYoga}.  On the contrary, causation is a viciously elusive thing to pin down.  But Chester, undaunted as always, was determined to give it a try. 

Chester first performed three basic tests for causality between the time series of spatial locations of the red dot and of the laser pointer, respectively.  These calculations were: the mutual information between time series, temporal precedence, and control for possible third variables.  The calculations all failed, due to insufficient observations of any object other than the dot.  Once appreciating that the red dot time series was really all he had to go on, Chester thought: well, what information did the red dot time series alone contain about the underlying dynamical system that produced it?  To investigate this question, Chester turned to time-delay embedding.

Time-delay embedding is a method of translating from a scalar set of measurements to a multivariate phase space that contains a geometric \textit{representation of} the (unobserved) dynamics that gave rise to the measurements~\cite{takens1981detecting,mane1981dynamical,eckmann1985ergodic}.  The geometric structure is called an attractor: a set of points in the space to which all nearby trajectories will tend to converge.  Since the theorem's publication in the 1980's, the procedure has seen a vast range of applications.  These include: characterizing chaos in dynamical systems~\cite{brown1991computing}, synthesizing musical tones~\cite{robel2001synthesizing}, analyzing dolphin echo-location~\cite{kremliovsky1998characterization,lainscsek2003characterization} and human speech~\cite{kokkinos2005nonlinear,johnson2005time,tishby1990dynamical}, using EKG recordings to identfffgdshjrify heart abnormalities~\cite{lainscsek2013electrocardiogram}, and using finger-tapping to predict Parkinson's Disease~\cite{lainscsek2012finger}.  To date, time delay embedding has not been employed to reconstruct the  dynamical system giving rise to the trajectory of an incorporeal red dot capable of traveling at the speed of light.  

Regarding this case of the red dot, Chester posed the question: \textit{does the resulting attractor in reconstructed phase space look like a laser pointer?}  After performing the  reconstruction, the answer appears to be no.  There exist various possible explanations, however, including the possibility that the red dot was in fact a hallucination brought on by COVID-19, the novel coronavirus disease caused by the virus SARS-Cov-2~\cite{cov}.  Moreover, the results of this study are inconclusive.  

Fortunately, Chester is not discouraged.  He has not missed a beajikllllllllll;'t.  Since concluding this study, he has attacked the renegade shadow of a housefly, chastized a billowing shower curtain, and now - glowering from under the bed - appears to be holding the entire household accountable for thunder.
\end{multicols}
\section{Data collection}
\begin{multicols}{2}

While the room in which the data collection occurred exists in three spatial dimensions, the analysis techniques employed in this paper require the use of scalar data sets.  Thus, Chester performed the analysis three times independently, for time series in the $x$, $y$, and $z$ directions, respectively.  For brevity, in this paper we show only the $x$ case, as it is representative of those for $y$ and $z$.

\begin{figure}[H]
\centering
\includegraphics[height=0.4\textheight]{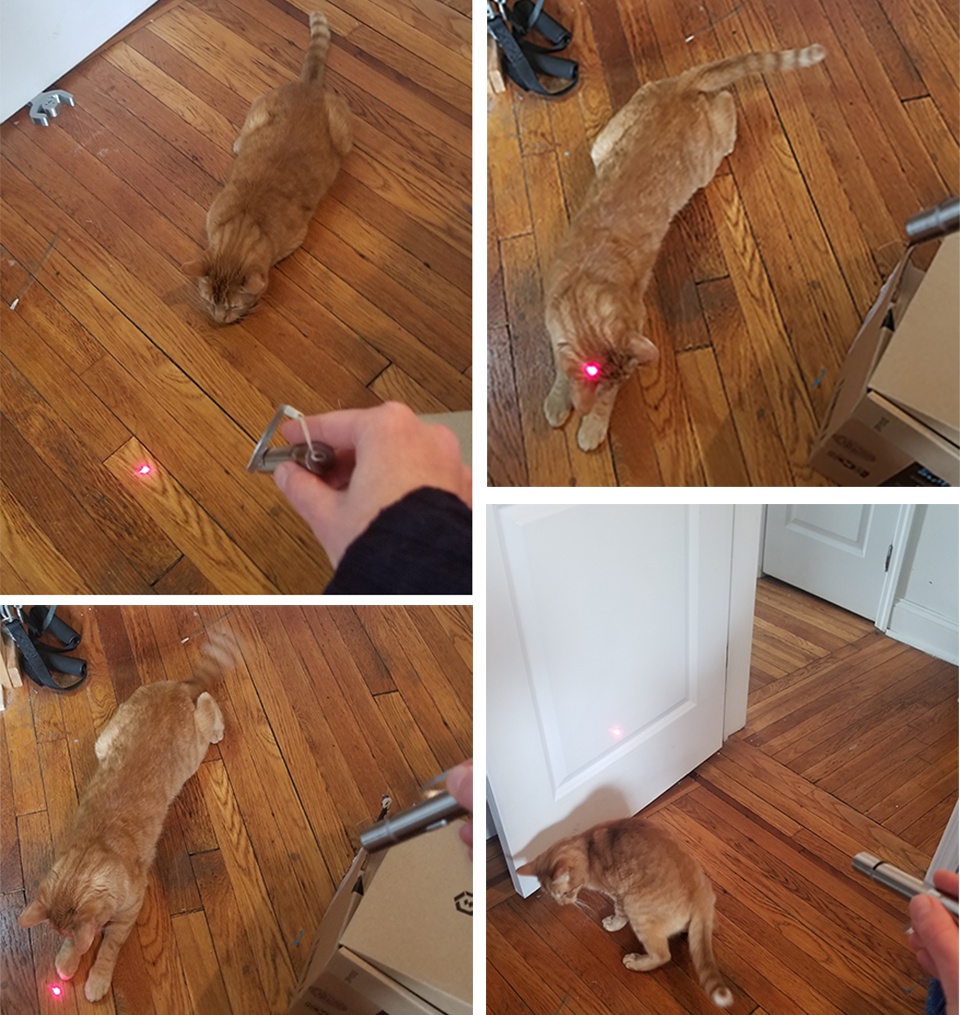}
\caption{\textbf{Chester gathers time series data} of the red dot and various objects and surfaces throughout the room, sampling at a step size of 0.1 s, the integration time of his eye.  He misses 12 points (e.g. top and bottom right panels), due to the incorporeal and light-speed properties of the dot.}
\label{fig:fig2}
\end{figure}

Chester first beheld the laser pointer in my hand at 9:04:17 am Eastern Standard Time on March 31, 2021, at the scalar location of $x=0.3$ in my husband Wayne's work room, where coordinate $x$ is defined as the direction from South ($x\equiv0.0$) to North ($x\equiv1.0$) walls.  He stared unblinking at the pointer for two seconds -- until 9:04:19, at which time his eyes darted away to scan the space nearby.  After two additional seconds (9:04:21), Chester located a red dot on the North wall.  Thereafter he sampled the red dot time series densely, using a step size of 0.1 s, the integration time of his eye  (Fig.~\ref{fig:fig2}).  The total baseline of the red dot time series was 130 s, at which time I grew bored.  During that time, Chester missed a total of only twelve temporal locations - a loss attributable not to any dearth of diligence on Chester's part, but rather to the noted incorporeal and light-speed properties of the dot (e.g. Fig.~\ref{fig:fig2}, top and bottom right panels).

\begin{figure}[H]
\centering
\includegraphics[width=0.8\textwidth]{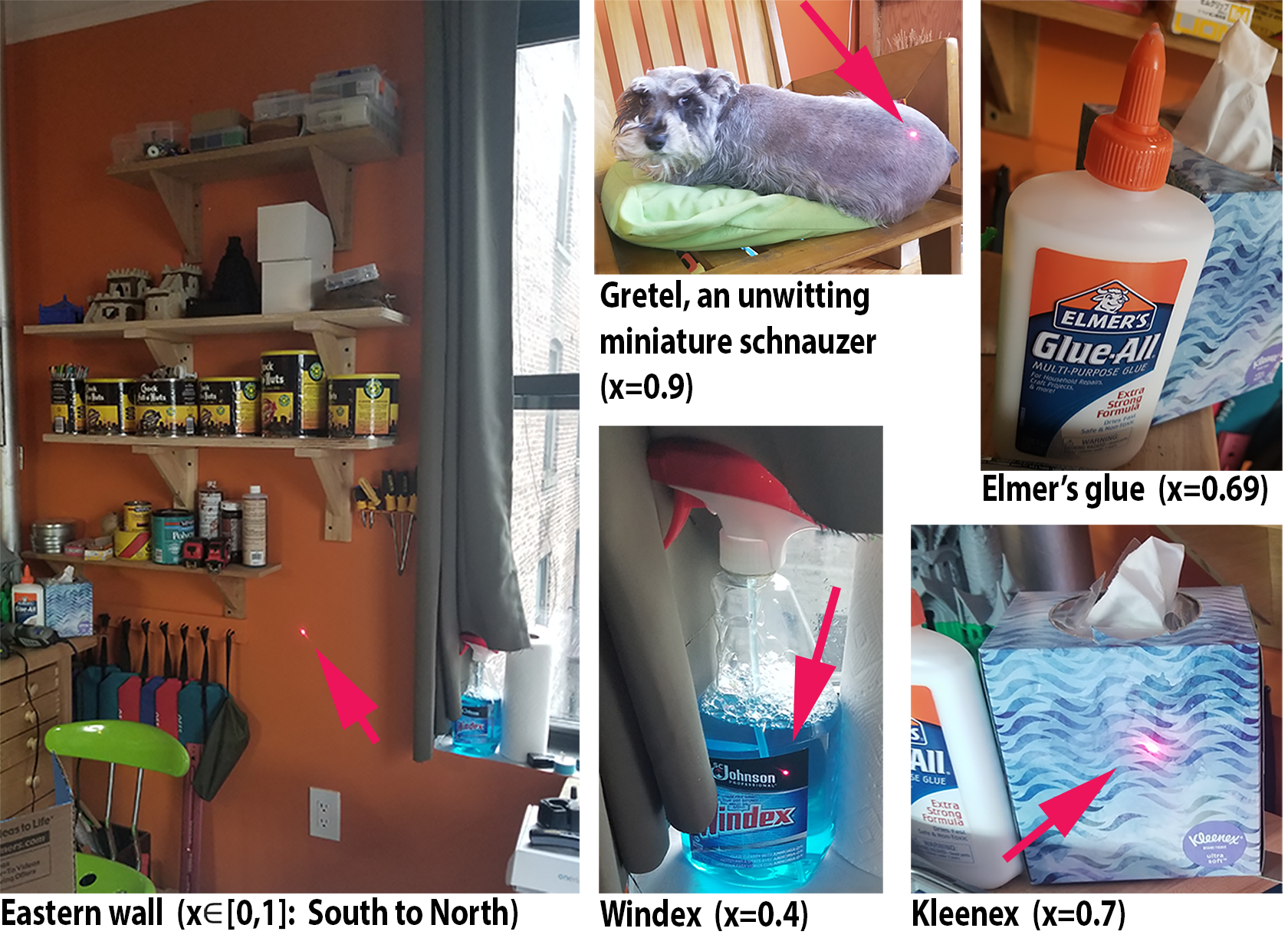}
\caption{\textbf{Objects throughout the room} displayed in scalar direction $x$: the direction from South to North in the room.  Red dot visits \textbf{Northernmost Wall} ($x \equiv 1.0$), \textbf{miniature schnauzer Gretel} ($x = 0.9$), a box of \textbf{Kleenex tissues} ($x = 0.7$), and a spray bottle of \textbf{Windex glass cleaner} ($x = 0.4$).  There exist multiple other objects in the room that the red dot does \textit{not} visit, all of which are represented by the tccccc \hspace{2cm} c \hspace{0.2cm} ccc \hspace{0.3cm}ime series of the \textbf{(unobserved) Elmer's glue} ($x = 0.69$).} 
\label{fig:fig3pics}
\end{figure}

Throughout the 130-second time series, the dot appeared upon various objects in the room.  Figure~\ref{fig:fig3pics} shows a representative sample, including the Eastern wall (left panel, spanning $x\in[0,1]$), Gretel - a napping miniature schnauzer (top middle, at $x=0.9$), a spray bottle of Windex (bottom middle, at $x=0.4$), and a box of Kleenex facial tissues (bottom right, at $x=0.7$).  Most of the objects in the room were in fact not visited by the red dot;  these unobserved -- or hidden -- variables are represented by the Elmer's Glue bottle (top right, at $x=0.69$).

Figure~\ref{fig:fig3timeSeries} shows Chester's observed time series: that of the red dot (top panel), laser pointer (second panel), four objects on which the red dot appeared, and one upon which it did not.  The data begin at $t\equiv0$ s, at which time the laser pointer time series (second panel) shows two seconds of observation.  A two-second gap then separates the end of the laser pointer time series and the beginning of the red dot time series ($t=4.0$ s), as Chester scanned for it.  From then onward, the red dot time series is dense,
\noindent
missing only twelve points total.

Panels 3 through 6 of Figure~\ref{fig:fig3timeSeries} show selected time series of objects that the dot visited, including the North wall (third panel; not to be confused with the East wall pictured in Figure~\ref{fig:fig3timeSeries}, left panel), Gretel the miniature schnauzer (fourth panel), Kleenex (fifth panel), and Windex (sixth panel).  Note the sparsity of these time series.  In addition, note that none of these objects moves in the $x$ dimension, with the one exception of Gretel, who - upon feeling a claw piercing her rear end - high-tailed it out of the room as quickly as possible.  

Finally, Panel 7 of Figure~\ref{fig:fig3timeSeries} shows Chester's time series sampling of the Elmer's Glue, an object upon which the red dot did not appear.  This time series is representative of all other objects upon which the red dot did not appear.  
\end{multicols}

\section{Preliminary tests to establish causality}
\begin{multicols}{2}

\subsection*{\textbf{Method}}
Before embarking on anything fancy, Chester first considered three basic requirements for establishing causality between two events: covariation, temporal precedence, and control for  third variables.

Now, covariance can be undefined if either variable does not have a well-defined mean.  Noting that the red dot trajectory appeared chaotic at times, and that the laser pointer trajectory contained a mere twenty data points (sampled at 0.1 s over two seconds), Chester prudently took neither to have a well-defined mean.  Thus, rather than covariance he opted to measure the mutual information~\cite{fano1961transmission} between the time series, as mutual information is always defined, and it has an established history as a test for causality~\cite{solo2008causality}.

The prescription for calculating mutual information goes as follows.  We call the time series of the red dot's trajectory $D = \{dot_m\}$ sampled over $m$ temporal locations, and the time series of the laser pointer's trajectory to be $P = \{pointer_n\}$ sampled over $n$ temporal locations.  Then we ask: How much information, in bits with logs to base 2, do we learn about Event $\{dot_m\}$ if we observed Event $\{pointer_n\}$, and vice versa?  Formally:
\begin{align*}
MI(dot_m,pointer_n) &= log\left[\frac{P_{DP}(d_m,p_n)}{P_D(dot_m)P_P(pointer_n)}\right]\\
&= MI(pointer_n,dot_m),
\end{align*}
\noindent
where $P_{DP}(d_m,p_n)$ is the joint probability distribution of events $dot_m$ and $pointer_n$, and the marginal distributions $P_D(dot_m)\equiv \sum_{pointer_n} P_{DP}(dot_m,pointer_n)$ and $P_P(pointer_n)\equiv \sum_{dot_m} P_{DP}(dot_m,pointer_n)$.
Then, the average over the elements is the average mutual information:
\end{multicols}
\begin{figure}[H]
\centering
\includegraphics[width=0.6\textwidth]{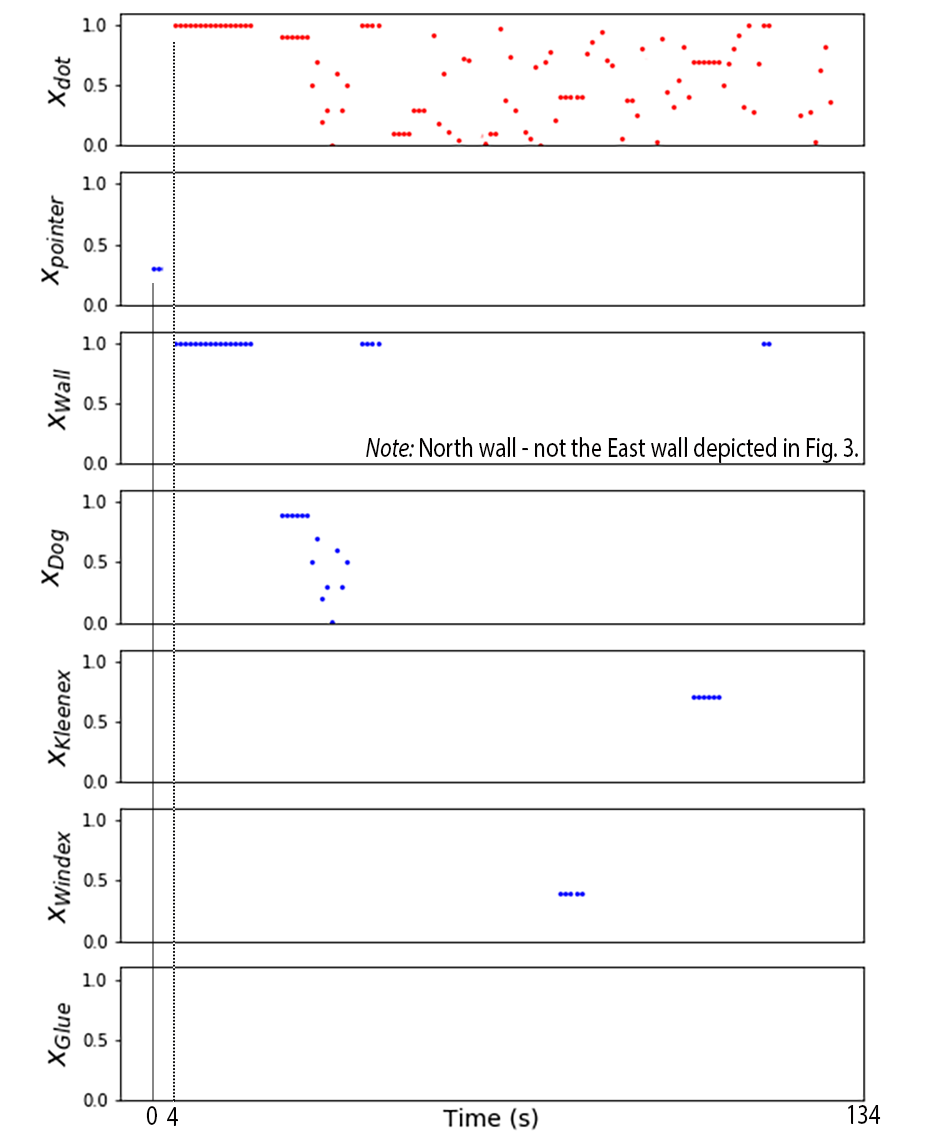}
\caption{\textbf{Chester's observed time series of selected objects in the room}.  \textit{From top}: \textbf{red dot}, \textbf{laser pointer}, \textbf{north wall}, \textbf{miniature schnauzer Gretel}, \textbf{Kleenex}, \textbf{Windex}, and \textbf{(unobserved) Elmer's glue}.  Note: Chester's sampling rate is 0.1s, but for brevity we plot at 1-s resolution with no loss of detail.  Chester first samples the laser pointer for two seconds before identifying the dot; grey vertical lines indicate the two-second gap between those time series.  Twelve points are missing between $t=4$ and 134 in the red dot time series, most occurring at times when laser appears to be varying chaotically in $x$.  Time series of Elmer's glue (\textit{bottom}) is representative of all other unsampled objects.} 
\label{fig:fig3timeSeries}
\end{figure}
\begin{multicols}{2}
\begin{align*}
AMI(D,P) &= \sum_{d_m,p_n} P_{DP}(dot_m,pointer_n)MI(dot_m,pointer_n).
\end{align*}
\noindent
The AMI measures the degree to which the two time series predict each other.

\subsection*{\textbf{Result}}

To machine precision, Chester's calculation of the average mutual information between the time series is zero.  This result is probably due to inadequate sampling.  While Chester made scrupulous observations of the flight of the red dot once he had spotted it ($m = 130$ s $\times 10$  samples/s $- 12 = 1288$), he noted only the first two seconds of the time series of the laser pointer's location ($n = 20$) and did not record the dot's time series during those initial two seconds.  

Temporal precedence also could not be established, as Chester had not bothered to check for the red dot at times prior to the appearance of the laser pointer.

Finally, Chester's monitoring for the role of a third variable was intermittent at best and nonexistent at worst (Panels 3 through 7 of Figure~\ref{fig:fig3timeSeries}).   

\end{multicols}
\section{Phase space reconstruction of the attractor underlying the red dot's flight}
\begin{multicols}{2}

\subsection*{\textbf{Method}}

Grudgingly accepting that the red dot time series was the only quality data he had to go on, Chester next turned to the more complicated - and admittedly in-parts-heuristic - method of time-delay embedding, to reconstruct the global phase space of the underlying dynamical system giving rise to the red dot's trajectory, and to thereby ascertain whether this dynamical system in any way resembled a laser pointer. 

Within the prescription of time-delay embedding, we assume that any observed scalar signal -- in this case, the $x(t)$ trajectory of our red dot -- arises from the mechanics of an underlying dynamical system, and that we can construct a geometric representation of that
\begin{figure}[H]
\centering
\includegraphics[width=0.75\textwidth]{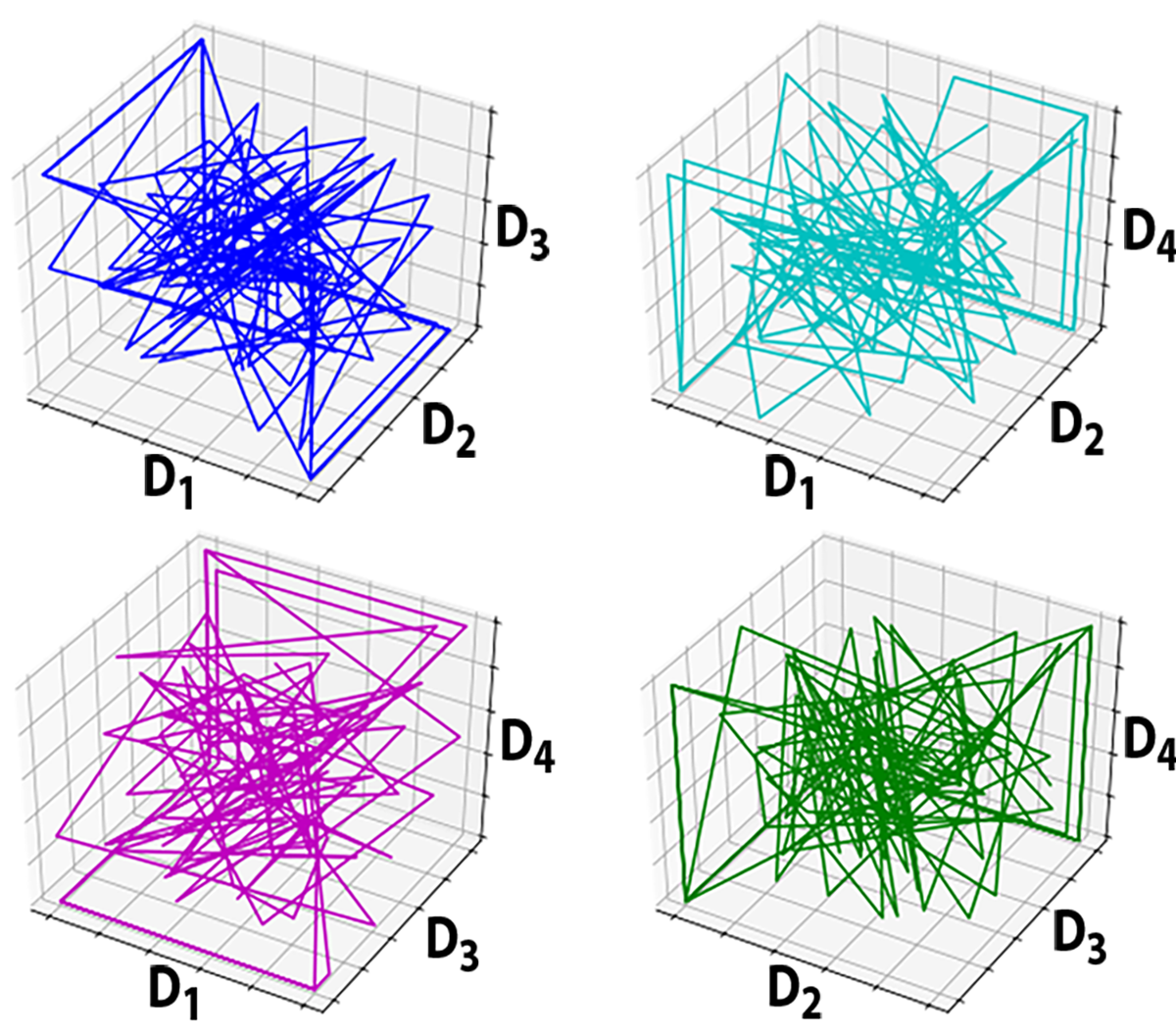}\\\vspace{0.1cm}
\includegraphics[width=0.8\textwidth]{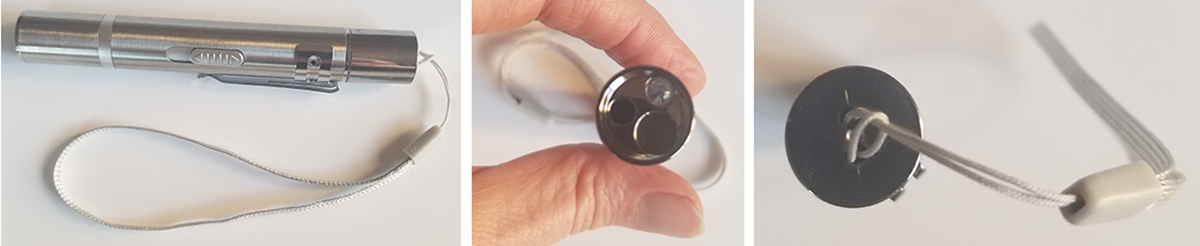}
\caption{\textbf{Attractors reconstructed from the $x(t)$ time series data set of the red dot (\textit{top}), and the laser pointer (\textit{bottom}).}  \textit{Top}: Four 3D visualizations of the reconstructed geometry of the dynamical system giving rise to the red dot's trajectory in the $x$ dimension.  Each of the four corresponds to a three-dimensional slice of the four-dimensional phase space, as $d=4$ was found to be the minimum dimension required for unfolding.  Here, the dimensions are labeled: $D1$, $D2$, $D3$, and $D4$.  These structures are representative of those $y(t)$ and $z(t)$ data sets (not shown).  \textit{Bottom}: Left to right: laser pointer in the x-z plane (representative of appearance in y-z plane), x-y plane at $z=0$, and x-y plane at $z=L$, where $L$ is the length of the laser pointer.  \textbf{No 3D slice of the geometry in reconstructed phase space resembles a laser pointer.}}
\label{fig:fig5}
\end{figure}
\noindent
system by means of the observations themselves.  The method has been used as a test for causality (e.g. Refs~\cite{krakovska2016testing,ye2015distinguishing,krakovska2018comparison}.  

Formally, the embedding theorem, attributed to Takens and Man\'{e}~\cite{takens1981detecting,sauer1991embedology,eckmann1985ergodic},  goes as follows.  We assume that there exists an unknown dynamical system $d\bm{x}/dt = \bm{H}(\bm{x}(t))$  and that we are observing some scalar projection $dot(\tikz\draw[black,fill=black] (0,0) circle (.5ex);)$ of some vector function of the dynamical variables $\bm{g}(\bm{x}(t))$.  Then the geometry of the unknown dynamics generating the measurements can be "unfolded" from those measurements $dot(\bm{g}(\bm{x}(t))$ in a space constructed of new vectors $\bm{y}(t)$ with components consisting of $dot(\tikz\draw[black,fill=black] (0,0) circle (.5ex);)$ applied to powers of $\bm{g}(\bm{x}(t)))$:  
\begin{align*}
  \bm{y}(t) &= [dot(\bm{x}(t)),dot(\bm{g}^{T_1}(\bm{x}(t))),...,dot(\bm{g}^{T_{d-1}}(\bm{x}(t)))],
\end{align*}
\noindent
where $d$ is the global dimension of the "reconstructed" phase space.  

This new set of vector $\bm{y}(t)$ describes an object in the space called an attractor.  An attractor is a point, or a set of points, in a phase space to which all nearby tattttyyytttttttttttttttttttytyttyttttttttttjifdlrajectories will tend to converge.  The vectors $\bm{y}(t)$ can be designed such that the new space preserves critical features of the original unknown dynamical system $d\bm{x}/dt = \bm{H}(\bm{x}(t))$; specifically, its deterministic - or temporal - behavior.  The embedding theorem applies to any smooth function of $dot(\tikz\draw[black,fill=black] (0,0) circle (.5ex);)$ and $\bm{g}(\bm{x})$~\cite{mindlin1991topological}.  

To $\bm{g}(\bm{x})$ we assign the operation that takes a vector $\bm{x}$ to that vector one time delay later.  So the $T_k^{th}$ power of $\bm{g}$ is: 
\begin{align*}
  \bm{g}^{T_k}(\bm{x}(t)) = \bm{x}(t_0 + (t+T_k)\tau_{sample}),
\end{align*}
\noindent
where $\tau_{sample}$ is the sampling rate.  Then, assuming that all delays are integer multiples of a common delay $\tau \equiv T_k\tau_{sample}$, the vectors $\bm{y}(t)$ become: 
\begin{equation} \label{eq:tdelay}
  \bm{y}(t) = [y(t), y(t + \tau), \dots, y(t + (d-1)\tau)]. 
\end{equation}
\noindent
It remains to choose the time delay $\tau$ and global embedding dimension $d$ that is sufficient to unfold the dynamics giving rise to the red dot trajectory (for these choices, see \textit{Appendix}), and then to plot the emergent vectors\footnote{For a detailed derivation of Equation~\ref{eq:tdelay} and a thorough pedagogical treatment of the technique, see Ref~\cite{abarbanel2012analysis}.}.

\subsection*{\textbf{R \hspace{4cm} vbbbbb \hspace{0.3cm} bb \hspace{0.5cm} besult}}
The data used for phase space reconstruction must be scalar quantities.  Thus Chester performed the analysis three independent times, for the $x$, $y$, and $z$ directions, respectively.  All three, if emerging from the same underlying mechanism, should yield the same dynamics.

For the analysis in $x$, $y$, and $z$ Chester found that the global embedding dimension required to unfold the dynamics underlying the observations was 4.  The stability of this value across the data sets was encouraging.  To visualize the emergent geomety in this space, Chester plotted every combination of three dimensions out of that four, for the $x$, $y$, and $z$ data sets independently.  The results for the $x$ set are representative of those for the other two, and are displayed in Figure~\ref{fig:fig5}.  None of the emergent attractors\footnote{Actually, we are not sure that these are attractors.  To establish this would require additional evidence such as Poincar\'{e} mapping~\cite{grassberger1983characterization}.} (top panel) resemble a laser pointer (bottom).  They aren't even silver\footnote{Do not bother trying to create a silver line plot via Matplotlib; it just looks grey.}.

Now, Chester's embedding procedure may have been flawed, for multiple reasons.  In addition to the heuristic choices for $\tau$ and $d$, the number of sampled points during the 130-second time series was smallish.  While some have had success with short data sets~\cite{ma2014detecting}, embedding tends to require time series of several tens of thousands of points~\cite{abarbanel2012analysis}.  Despite the ambiguous significance of the attractors pictured in Figure~\ref{fig:fig5} they are in fact quite intriguing and merit discussion (see \textit{Discussion}).  
\end{multicols}
\section{Or maybe it's COVID-19}
\begin{multicols}{2}
\begin{figure}[H]
\centering
\includegraphics[width=0.7\textwidth]{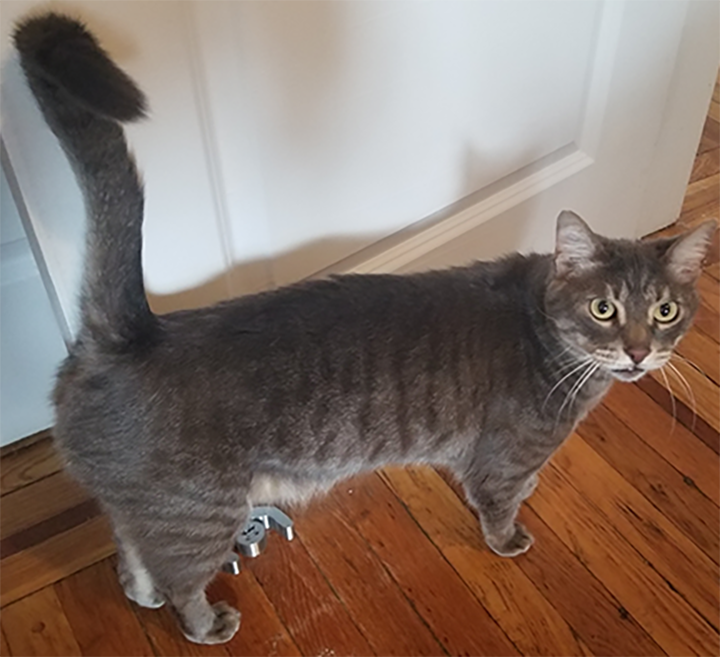}
\caption{\textbf{Mad Dog Lapynski, Chester's brother.}  Wandered in during the data collection.}
\label{fig:fig6}
\end{figure}
While Chester recalls the dot/pointer correlation from past occasions, he allowed that this particular instance \textit{might} be pegged on COVID-19 -- because, well, at the moment most anything~\cite{covid1} might be pegged on COVID-19.  That is, could the red dot be a hallucination?

\begin{figure}[H]
\centering
\includegraphics[width=0.87\textwidth]{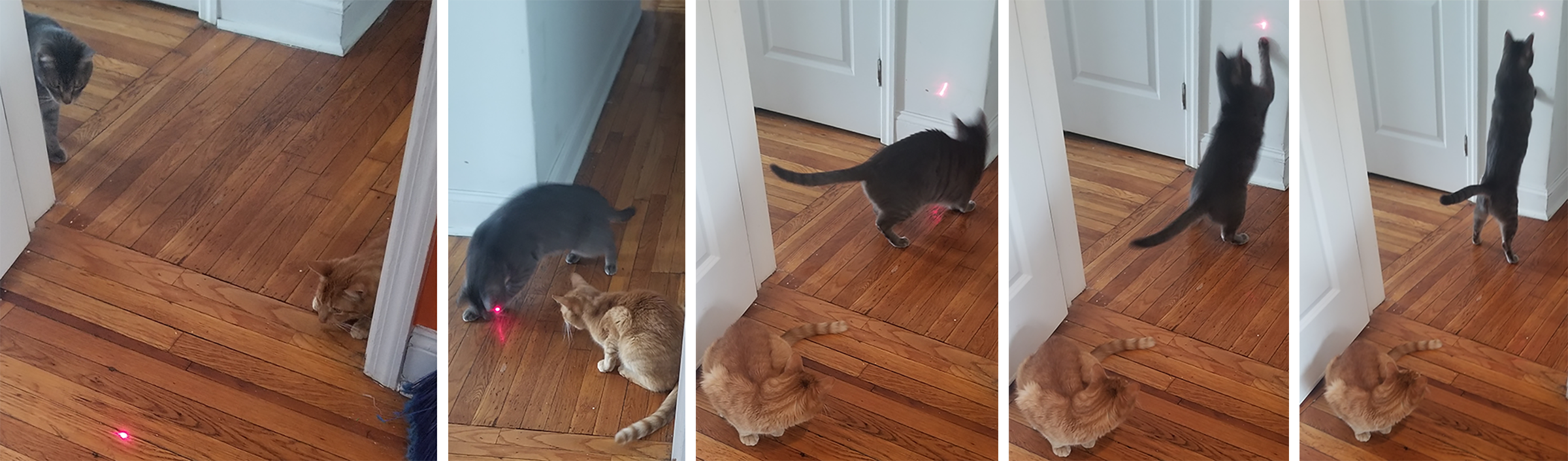}
\caption{\textbf{Mad Dog Lapynski also sees the red dot}, an independent check that renders the COVID-19 hypothesis unlikely.}
\label{fig:fig7}
\end{figure}

Fortunately, as Chester pondered this possibility, his younger brother Mad Dog Lapynski (Figure~\ref{fig:fig6}) wandered onto the scene.  Mad Dog Lapynski also noted the red dot (Figure~\ref{fig:fig7}), inadvertently establishing an independent verification of its existence\footnote{Unless the brothers are displaying identical symptoms.}.  

Mad Dog Lapynski opted not to perform his own analysis of the data.  He only stayed long enough to comment snidely that Chester's study neglected to consider topological causality~\cite{harnack2017topological}, and to smugly decline our offer of co-authorship.
\end{multicols}
\begin{multicols}{2}
\end{multicols}
\section{DISCUSSION}
\begin{multicols}{2}

While the strategies employed in this paper have yielded inconclusive results, they by no means represent an exhaustive examination of tests for causality.  Other lines of attack exist in the literature, for example, topological causality~\cite{harnack2017topological}, as noted by Mad Dog Lapynski.  Nevertheless, this study is noteworthy as the first dynamical systems approach to probing the confounding nature of incorporeal light-speed red dots.  

In addition, while the significance of the attractors in Chester's reconstructed phase space is unclear, the results do carry intriguing implications that might not be readily obvious to the reader.  Allow us to elucidate.  

FTake, for example, the structure that emerges in the $D_1$-$D_3$-$D_4$ slice of the four-dimensional phase space (Figure~\ref{fig:fig5}, middle left).  Thisf particular viewing angfffle reveals the startfffffffffffling property offfffffffffffffffffffffffffffgrtoity[jlu;mj9jgffggggggggffffffffffffffffffffffffffffffffffffffffffff ffffffffffffffffffffffffffffffffffffffffffffffffffffffffffffffffffffffffffffff 
ffffffffffffffffffffffffffffffffffffffffffffffffffffffffffffffffffffffffffffff 
ffffffffffffffffffffffffffffffffffffffffffffffffffffffffffffffffffffffffffffff 
ffffffffffffffffffffffffffffffffffffffffffffffffffffffffffffffffffffffffffffff 
ffffffffffffffffffffffffffffffffffffffffffffffffffffffffffffffffffffffffffffff 
ffffffffffffffffffffffffffffffffffffffffffffffffffffffffffffffffffffffffffffff 
ffffffffffffffffffffffffffffffffffffffffffffffffffffffffffffffffffffffffffffff 
ffffffffffffffffffffffffffffffffffffffffffffffffffffffffffffffffffffffffffffff 
ffffffffffffffffffffffffffffffffffffffffffffffffffffffffffffffffffffffffffffff 
ffffffffffffffffffffffffffffffffffffffffffffffffffffffffffffffffffffffffffffff 
ffffffffffffffffffffffffffffffffffffffffffffffffffffffffffffffffffffffffffffff 
ffffffffffffffffffffffffffffffffffffffffffffffffffffffffffffffffffffffffffffff 
ffffffffffffffffffffffffffffffffffffffffffffffffffffffffffffffffffffffffffffff 
ffffffffffffffffffffffffffffffffffffffffffffffffffffffffffffffffffffffffffffff 
ffffffffffffffffffffffffffffffffffffffffffffffffffffffffffffffffffffffffffffff 
ffffffffffffffffffffffffffffffffffffffffffffffffffffffffffffffffffffffffffffff 
ffffffffffffffffffffffffffffffffffffffffffffffffffffffffffffffffffffffffffffff 
ffffffffffffffffffffffffffffffffffffffffffffffffffffffffffffffffffffffffffffff 
ffffffffffffffffffffffffffffffffffffffffffffffffffffffffffffffffffffffffffffff 
ffffffffffffffffffffffffffffffffffffffffffffffffffffffffffffffffffffffffffffff 
ffffffffffffffffffffffffffffffffffffffffffffffffffffffffffffffffffffffffffffff 
ffffffffffffffffffffffffffffffffffffffffffffffffffffffffffffffffffffffffffffff 
ffffffffffffffffffffffffffffffffffffffffffffffffffffffffffffffffffffffffffffff 
ffffffffffffffffffffffffffffffffffffffffffffffffffffffffffffffffffffffffffffff 
ffffffffffffffffffffffffffffffffffffffffffffffffffffffffffffffffffffffffffffff 
ffffffffffffffffffffffffffffffffffffffffffffffffffffffffffffffffffffffffffffff 
ffffffffffffffffffffffffffffffffffffffffffffffffffffffffffffffffffffffffffffff 
ffffffffffffffffffffffffffffffffffffffffffffffffffffffffffffffffffffffffffffff 
ffffffffffffffffffffffffffffffffffffffffffffffffffffffffffffffffffffffffffffff 
ffffffffffffffffffffffffffffffffffffffffffffffffffffffffffffffffffffffffffffff 
ffffffffffffffffffffffffffffffffffffffffffffffffffffffffffffffffffffffffffffff 
ffffffffffffffffffffffffffffffffffffffffffffffffffffffffffffffffffffffffffffff 
ffffffffffffffffffffffffffffffffffffffffffffffffffffffffffffffffffffffffffffff 
ffffffffffffffffffffffffffffffffffffffffffffffffffffffffffffffffffffffffffffff 
ffffffffffffffffffffffffffffffffffffffffffffffffffffffffffffffffffffffffffffff 
ffffffffffffffffffffffffffffffffffffffffffffffffffffffffffffffffffffffffffffff 
ffffffffffffffffffffffffffffffffffffffffffffffffffffffffffffffffffffffffffffff 
ffffffffffffffffffffffffffffffffffffffffffffffffffffffffffffffffffffffffffffff 
ffffffffffffffffffffffffffffffffffffffffffffffffffffffffffffffffffffffffffffff 
ffffffffffffffffffffffffffffffffffffffffffffffffffffffffffffffffffffffffffffff 
ffffffffffffffffffffffffffffffffffffffffffffffffffffffffffffffffffffffffffffff 
ffffffffffffffffffffffffffffffffffffffffffffffffffffffffffffffffffffffffffffff 
ffffffffffffffffffffffffffffffffffffffffffffffffffffffffffffffffffffffffffffff 
ffffffffffffffffffffffffffffffffffffffffffffffffffffffffffffffffffffffffffffff 
ffffffffffffffffffffffffffffffffffffffffffffffffffffffffffffffffffffffffffffff 
ffffffffffffffffffffffffffffffffffffffffffffffffffffffffffffffffffffffffffffff 
ffffffffffffffffffffffffffffffffffffffffffffffffffffffffffffffffffffffffffffff 
ffffffffffffffffffffffffffffffffffffffffffffffffffffffffffffffffffffffffffffff 
ffffffffffffffffffffffffffffffffffffffffffffffffffffffffffffffffffffffffffffff 
ffffffffffffffffffffffffffffffffffffffffffffffffffffffffffffffffffffffffffffff 
ffffffffffffffffffffffffffffffffffffffffffffffffffffffffffffffffffffffffffffff 
fffscinating and merits further investigation.

\end{multicols}

\section{ACKNOWLEDGMENTS}
\begin{multicols}{2}
E.A. thanks her attorney overseeing the action filed by her co-author concerning the appropriate assignment of first authorship of this manuscript.  (Her case is that she did all of the write-up; C. argues that he did all of the work and \textit{some} of the write-up.)  E.A. also thanks BAND-AID \textsuperscript{\textregistered} regarding the multiple injuries incurred by her forearms throughout the litigation thus far.  C., as usual, thanks no one.
\end{multicols}

\section*{Appendix}

Chester's choices for the values of time delay $\tau$ and global embedding dimension $d$ for phase space reconstruction were rather heuristic, as they have been in all the literature cited in this paper. 

To choose $\tau$ there exists no single formal prescription.  The delay should be sufficiently long to render the coordinates essentially independent, so that they form an orthonormal basis that spans the phase space.  Too short a delay will not permit sufficient new information to have been generated in the interim, while too long a delay will erase evidence of causality, assuming that there was any to begin with.  To strike a balance between these extremes, Chester plotted the average mutual information between coordinates $AMI(x(t)\mid x(t+\tau))$ as a function of $\tau$, and chose the value of $\tau$ that yielded the first minimum of this distribution~\cite{fraser1986independent,fraser1989information}.  This strategy has been used successfully for time-delay embedding of the chaotic H\'enon map~\cite{brown1991computing}, chaotic Lorenz-63 system~\cite{brown1991computing}, and an audio tape taken inside an aviary of ornery cowbirds~\cite{armstrong2018colonel}.  (Note: here we are considering the mutual information between successive points in the red-dot time series, and not between the dot time series and laser pointer time series, as was the case in \textit{Preliminary tests}.)

Next, the choice for $d$ performs the unfolding.  There exists no single preferred methodology for this calculation either.  Chester followed the prescription of minimizing the percentage of \lq\lq false nearest neighbors\rq\rq~\cite{kennel1992determining} in a time series.  On a trajectory, points may appear sequential for one of two reasons: 1) the dynamics truly renders them sequential, or 2) errors in the sequence were acquired via the projection from the $\bm{x}(t)$ space to the scalar $h(\bm{g}(\bm{x}(t)))$ space -- and so these sequential points are "false neighbors".  If a sufficient number of components $d$ is used to construct the vectors $\bm{y}(t)$ that define our new phase space, then the tangles can be unfolded - i.e. the imposter neighbors eliminated. 
\bibliographystyle{unsrt}
\nocite{*}
\bibliography{refs}

 


\end{document}